\documentclass[twocolumn,superscriptaddress,floatfix]{revtex4-2}

\usepackage{graphicx}
\usepackage[para]{threeparttable}
\usepackage{amsmath}
\usepackage{lineno}
\usepackage{booktabs}
\usepackage[colorlinks=true,linkcolor=blue]{hyperref}

\begin{document}

\title[]{Probabilistic neural networks for improved analyses with phenomenological models} 

\author{C.~H.~\surname{Kim}}
\affiliation{Department of Physics, Sungkyunkwan University, Suwon 16419, Republic of Korea}

\author{K.~Y.~\surname{Chae}}
\email{kchae@skku.edu}
\thanks{Fax: +82-31-290-7055}
\affiliation{Department of Physics, Sungkyunkwan University, Suwon 16419, Republic of Korea}

\author{M.~S.~\surname{Smith}}
\affiliation{Physics Division, Oak Ridge National Laboratory, Oak Ridge, Tennessee 37831, USA}

\author{D.~W.~\surname{Bardayan}}
\affiliation{Department of Physics and Astronomy, University of Notre Dame, Notre Dame, Indiana 46556, USA}

\author{C.~R.~\surname{Brune}}
\affiliation{Department of Physics and Astronomy, Ohio University, Athens, Ohio 45701, USA}

\author{R.~J.~\surname{deBoer}}
\affiliation{Department of Physics and Astronomy, University of Notre Dame, Notre Dame, Indiana 46556, USA}

\author{D.~\surname{Lu}}
\affiliation{Computational Sciences and Engineering Division, Oak Ridge National Laboratory, Oak Ridge, Tennessee 37831, USA}

\author{D.~\surname{Odell}}
\affiliation{Department of Physics and Astronomy, Ohio University, Athens, Ohio 45701, USA}

\begin{abstract}

We present a method for measurement analyses based on probabilistic deep neural networks that provide several advantages over conventional analyses with phenomenological models. These include the prediction of objects directly from data, the rapid generation of statistically robust uncertainties, and the ability to bypass some parameters that may induce ambiguities and complications in data analysis. As deep learning methods make predictions through ``black boxes,'' the uncertainty quantification is typically challenging. We use a probabilistic framework that provides thorough uncertainty quantification and is straightforward to follow in practice. With the Transformer, we demonstrate the current method for predicting nuclear resonance parameters from scattering data using the phenomenological $R$-matrix model.

\end{abstract}

\maketitle

In modern physics, phenomenological models are routinely used to understand physical systems by fitting their predictions to measured data and extracting ``best fit'' values for the model parameters \cite{C.Cutler1994, P.Moller2016, H.Wiltsche2020, D.Everett2021}. To determine their numerous parameters, each of which has its own particular contribution to observables, $\chi^2$ minimization or Bayesian inference are typically used \cite{R.DAgostino1986, U.Toussaint2011}. In some cases, however, the quantification of uncertainties is not straightforward and their calibrations are questionable \cite{A.Dawid1982, D.Smith2007}. Additionally, models may have some parameters that do not correspond to observable features in measurements, but rather induce complications and ambiguities during the interpretation of measurement results \cite{P.Descouvemont2010, R.deBoer2017}.

Such challenges motivate the exploration of new data analysis approaches. Deep learning, with its ever-increasing capabilities, is currently one of the most promising alternatives to mitigate the difficulties of conventional analysis methods. Deep learning approaches utilize high capacity deep neural networks (DNNs) with the flexibility to approximate a wide variety of models \cite{L.Wright2022, F.Ashtiani2022, J.Jumper2021, M.Raissi2019}, making their adoption the norm for data analyses in some areas of physics \cite{C.Kim2023, C.Kim2023-2, H.Gabbard2022, A.Boehnlein2022}. While the inherent ``black box'' character of standard DNNs usually impedes the assessment of uncertainties, incorporating probabilistic frameworks in DNNs enables uncertainty quantification using stochastic features \cite{M.Abdar2021, L.Jospin2022, K.Murphy2022}, thereby broadening their utility for physics research.

\begin{figure}[!ht]
\centering
\includegraphics[width=0.47\textwidth]{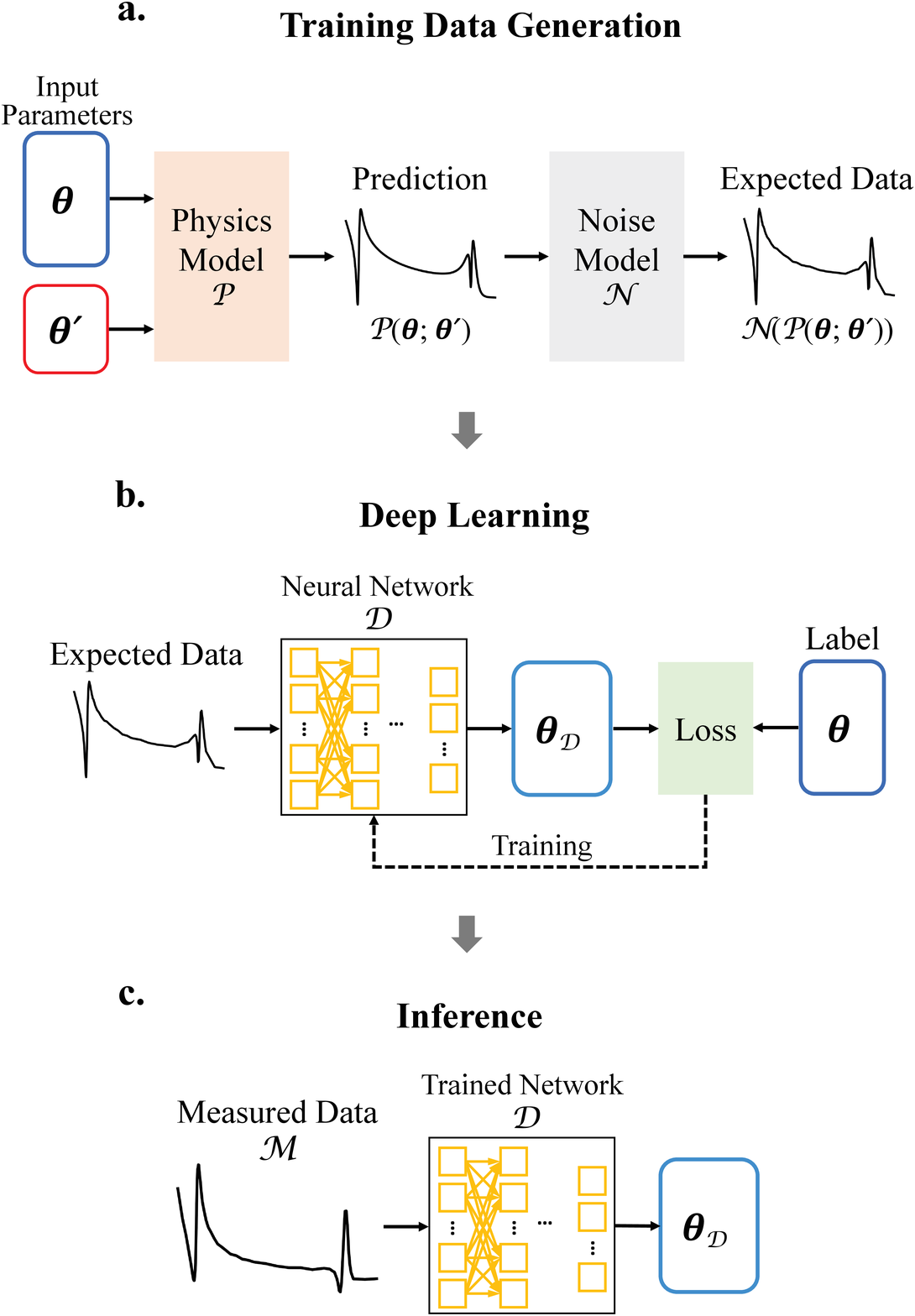}
\caption{Workflow of our method. $\textbf{a}$, Training data generation using the physics model with two subsets of parameters $\boldsymbol{\theta}$ and $\boldsymbol{\theta}^\prime$. $\textbf{b}$, Deep learning modeling with training data labeled only by the parameters of interest, $\boldsymbol{\theta}$, to yield predictions $\boldsymbol{\theta}_\mathcal{D}$. $\textbf{c}$, Inference from the measurement. The trained model requires only measurement data to extract parameters at the inference stage.}
\label{fig1}
\end{figure}

This study introduces a general purpose, straightforward method based on a probabilistic deep learning framework known as deep ensembles that offers numerous advantages over data analyses with phenomenological models. We demonstrate that our trained deep learning model has the capability to reliably and rapidly extract features directly from data with statistically calibrated quantification of uncertainties. Furthermore, by suitable training approaches, model parameters that induce ambiguities in data interpretations can in some cases be bypassed with our widely applicable method.

Fig.~\ref{fig1} shows a sketch of our method to determine parameters of phenomenological models from measurement data. First, as shown in Fig.~\ref{fig1}$\textbf{a}$, training data can be obtained by running a phenomenological model $\mathcal{P}$ with various input combinations of parameters of interest $\boldsymbol{\theta}$ and, if present, any other extraneous parameters $\boldsymbol{\theta}^\prime$ required by the model that may introduce ambiguities in the analysis. The outputs can be processed with a noise function $\mathcal{N}$ to simulate the noise present in experimental measurements, generating the ``expected data.'' Then, as shown in Fig.~\ref{fig1}$\textbf{b}$, a deep neural network $\mathcal{D}$ is trained on this data using the expected data as inputs and using $\boldsymbol{\theta}$ as the data labels. The key to eliminate extraneous parameters $\boldsymbol{\theta}^\prime$ is to label the training data with $\it{only}$ the parameters of interest $\boldsymbol{\theta}$, so that the deep learning model can extract these from the data without any knowledge of the unwanted parameters $\boldsymbol{\theta}^\prime$. By training over data generated with a wide range of these parameter values, we ensure that the deep learning model works on data with any parameter values. After training, the model is used to predict the objectives directly from measured data $\mathcal{M}$, where $\boldsymbol{\theta}_\mathcal{D}$ represents the prediction (Fig.~\ref{fig1}$\textbf{c}$). 

The use of probabilistic neural networks in our method can give particular improvements in the measurement analysis with thorough uncertainty quantification. Bayesian neural networks with Monte Carlo samplings can be suitable for this purpose. However, they are normally difficult to implement for non-experts and, because of their computational requirements \cite{L.Jospin2022, K.Murphy2022}, they are limited to very small networks -- which are incapable of modeling complex physics phenomena. Here, we introduce probabilistic frameworks known as deep ensembles which are easy to implement and applicable to large networks \cite{B.Lakshminarayanan2017}. Deep ensembles have been validated as a straightforward and highly effective method, featuring the use of multiple point-estimate models which are trained, using a negative log-likelihood for the loss function, from random initial model parameters \cite{Y.Ovadia2019, L.Jospin2022, M.Abdar2021}. The model prediction $\boldsymbol{y}$ on given input data $\boldsymbol{x}$ can be obtained from the predictions of these multiple models:
\begin{linenomath}
\begin{equation}
p(\boldsymbol{y} \!\mid\! \boldsymbol{x}) = \frac{1}{N_{\text{model}}}\sum_{n=1}^{N_{\text{model}}} p(\boldsymbol{y} \!\mid\! \boldsymbol{x}, \boldsymbol{\theta}_n), 
\label{eq1}
\end{equation}
\end{linenomath}
where $\boldsymbol{\theta}_n$ represents parameters of the $n$th model. As shown in Eq.~\ref{eq1}, the predictive distribution $p(\boldsymbol{y} \!\mid\! \boldsymbol{x})$ is the average of the likelihoods $p(\boldsymbol{y} \!\mid\! \boldsymbol{x}, \boldsymbol{\theta}_n)$ from the models. In classification, then, the predictive distribution will be the average of the softmax outputs from the models. In regression, each model should output the information of likelihoods. If one assumes the likelihood follows Gaussian, the model should have two outputs for each prediction, the mean and sigma. From these, one can obtain the likelihoods, and therefore the predictive distribution.

The idea of deep ensembles can be interpreted as another approach for Bayesian deep learning using variational inference \cite{L.Jospin2022, A.Wilson2020}. In practice, it can be done by simply training multiple models with different initial parameters. From such neural networks, we can expect models capable of rapidly generating statistically validated probabilistic predictions for both discrete and continuous quantities with minimal computational requirements.

\begin{table}[!t]
\resizebox{0.47\textwidth}{!}{
\def\arraystretch{0.5}
\begin{threeparttable}
\caption{Training data distribution and test results of our model. Model performance is presented with accuracy (for $J$ and $\pi$) and mean errors (for $E$ and $\Gamma$). Median values of predictive distributions are taken to calculate the errors. In total, 5 BGPs are included.}\label{tab1}
\begin{tabular}{ccc}
\toprule
Parameters & Data Distribution & Model Performance \\
\midrule
$J^{\pi}_{1}$ & $\frac{1}{2}^{+}$ (Fixed) & $J$: n/a, $\pi$: n/a  \\ \\
$J^{\pi}_{2}$ & $l \leq 2$ & $J$: 98.7 $\%$, $\pi$: 100.0 $\%$ \\ \\
$J^{\pi}_{3}$ & $l = 2$ & $J$: 99.0 $\%$, $\pi$: n/a \\ \\
$E_{1}$ (MeV) & Normal(0.425, 0.003) & 0.00014 \\ \\
$E_{2}$ (MeV) & Normal(1.565, 0.010)\tnote{3} & 0.00124 \\ \\
$E_{3}$ (MeV) & Normal(1.610, 0.006)\tnote{3} & 0.00059 \\ \\
$\Gamma_{1}$ (keV) & Normal(30, 3) & 0.20 \\ \\
$\Gamma_{2}$ (keV) & Normal(50, 15) & 1.98 \\ \\
$\Gamma_{3}$ (keV) & Normal(60, 8) & 1.32 \\
\midrule
ANC (fm$^{-1/2}$)\tnote{1} & Normal(1.81, 0.07) & n/a \\ \\
$a_c$ (fm) & 3 - 8 & n/a \\ \\
5 BGPs [$J^{\pi}$] & $l \leq 2$ & n/a \\ \\
5 BGPs [$E$] & Uniform(3, 10) & n/a \\ \\
5 BGPs [$\Gamma$] & Uniform(0, $\Gamma_W$)\tnote{2} & n/a \\
\bottomrule
\end{tabular}
\begin{tablenotes}
\linespread{1.2} \footnotesize
\item [1] Asymptotic normalization coefficient (ANC) for the subthreshold state is also included in the $\textit{R}$-matrix calculation but not as one of the objects to simplify the demonstration. The range is taken from the previous study \cite{S.Artemov2003, R.Azuma2010}.
\item [2] $\Gamma_W$ is the proton width calculated from $\Gamma_W = 2P\gamma_W^2 = 2P(3\hbar^2/2\mu a_c^2)$, where $P$ is the penetrability, $\gamma_W^2$ is the Wigner limit, and $\mu$ is the reduced mass.
\item [3] We set $E$ of the second and third states to be separated more than 10 keV as the energy bin sizes of the measurement data in this region are $\geq$10 keV.
\end{tablenotes}
\end{threeparttable}}
\end{table}

We demonstrate our method on the phenomenological $R$-matrix model that is widely used in nuclear physics \cite{A.Lane1958, H.Fynbo2005, P.Descouvemont2010, R.Azuma2010, E.Adelberger2011, R.deBoer2017, A.Tumino2018, J.Bishop2022}. This model is based on separating the particle interaction space into internal and external regions whose boundary is defined by a parameter known as the channel radius. Complex many-body nuclear interactions are present in the internal region and are characterized by parameters of nuclear resonant states. From the $\textit{R}$-matrix that contains such parameters, expected observational data such as cross sections can be calculated. The properties of nuclear states are determined by comparing the expected and experimental data. There are, however, difficulties in data analyses using the $R$-matrix model -- primary among them being the existence of the channel radius and background poles included to compensate the truncated higher lying states in $R$-matrix calculations \cite{R.deBoer2017}. These may induce complications in analyses due to the rather arbitrary choices of these parameters, as there are no strict physics rules to guide their values \cite{A.Lane1958, P.Descouvemont2010, R.deBoer2017}.

One of the major applications of the phenomenological $\textit{R}$-matrix is determining resonance parameters (e.g., spins $J$, parities $\pi$, energies $E$, widths $\Gamma$) in nuclear reaction data. Our deep learning model was built to find patterns of resonance parameters in reaction cross section spectra. Specifically, our goal was to extract the parameters of three resonance states from the well-known $^{12}$C+$p$ elastic scattering measurement of ref. \cite{H.Meyer1976}. In our approach, we also bypassed the need to determine the channel radius and background poles, eliminating ambiguities that these can introduce.

The training data for this task was prepared with an $R$-matrix code, $AZURE2$ \cite{R.Azuma2010}, via the following specific process. We sampled possible sets of resonance parameters, along with random values of the channel radius ($a_c$) and random configurations of background poles (BGPs). We assumed a typical situation where previous studies have determined some of the resonance parameters, to demonstrate that any case can be handled. Table~\ref{tab1} shows the training data distributions of $R$-matrix model parameters. Here, $l$ is the relative orbital angular momentum of the particle pair. Possible spins and parities with the given $l$ were uniformly sampled: e.g., $J=0.5-2.5$ and $\pi=+, -$ for $l \leq 2$. We set the range of the channel radius to 3-8 fm that contains typical values being used in traditional $\textit{R}$-matrix calculations. For $E$ and $\Gamma$, we used Gaussian distributions as if some previous studies had constrained them. A level for each possible spin-parity was included as a background pole. The location and width of these levels were randomly determined within the given ranges in Table~\ref{tab1}. 

\begin{figure}[!t]%
\centering
\includegraphics[width=0.47\textwidth]{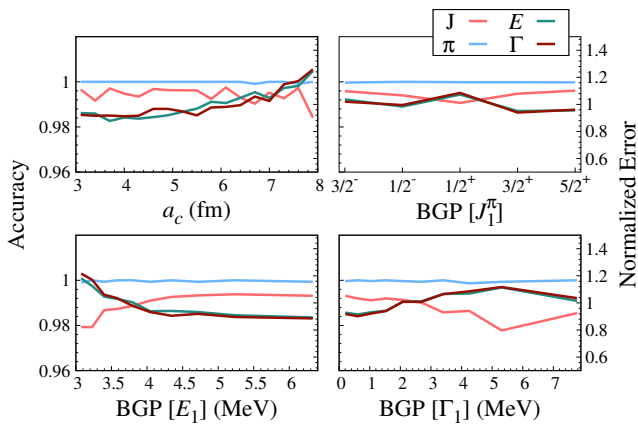}
\caption{Errors of the model predictions for cross sections calculated with various unobservables. Res 1, 2, and 3 represent the first, second, and third resonances. The figures of the first background pole are given as representations.}\label{fig2}
\end{figure} 

We calculated cross section spectra corresponding to each set of parameters using the $\textit{R}$-matrix theory \cite{R.Azuma2010}. As the calculated spectra do not contain experimental noise, we added random noise in the spectra assuming Gaussian noises, where the standard deviations were obtained from the measurement errors given by ref. \cite{H.Meyer1976}. We set the log of the differential cross sections as the data inputs to the model, and we set the data labels to be the corresponding (encoded) values of parameters $J$, $\pi$, $\text{log}(E)$, and $\text{log}(\Gamma)$ of three resonance states.

We used the Transformer architecture as the foundation for our model. The Transformers are currently one of the most effective deep learning architectures used in a wide variety of applications \cite{A.Vaswani2017, J.Devlin2019, S.Khan2022, H.Zhou2021}. They are based on an encoder-decoder structure and feature an ``attention'' mechanism to handle sequence data. We modified Transformers to work with cross section spectra at multiple angles. Because of the different nature of our four parameters $J$, $\pi$, $E$, and $\Gamma$, we split the architecture output into four branches where two of them predict discontinuous values (classification) for $J$ and $\pi$, and the others predict continuous values (regression) for $\text{log}(E)$ and $\text{log}(\Gamma)$. We chose reasonable values for model hyperparameters based on several empirical test runs, but did not fine-tune their values.

\begin{figure}[!t]%
\centering
\includegraphics[width=0.47\textwidth]{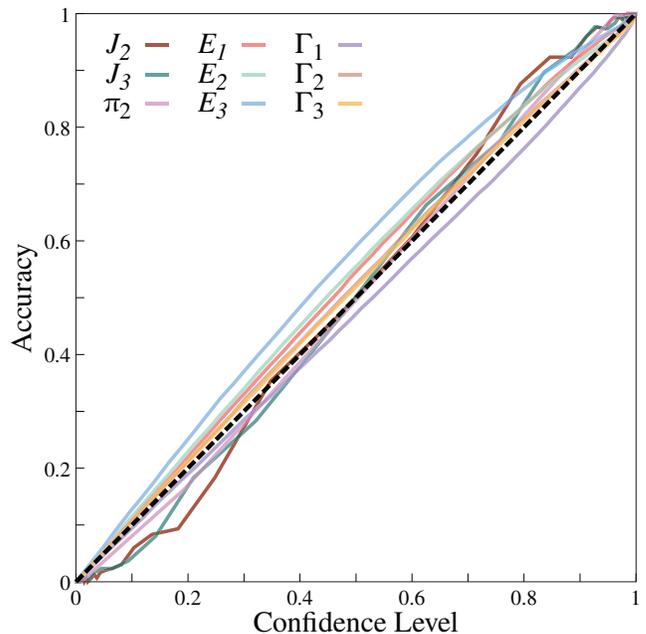}
\caption{Reliability diagram of the current model. Each curve shows the estimated accuracy (fraction of true data points included in the confidence range) as a function of the corresponding confidence level. It is the standard method to evaluate the calibration of uncertainty.}\label{fig3}
\end{figure}

\begin{figure}[!t]%
\centering
\includegraphics[width=0.47\textwidth]{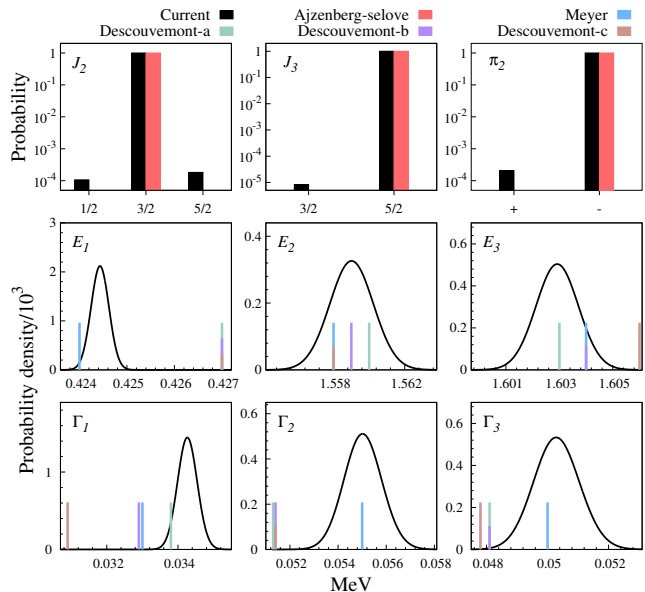}
\caption{Predictions of the resonance parameters from the measurement data. $J$ and $\pi$ in the literature and $E$ and $\Gamma$ presented by the previous fitting results are also shown -- Ajzenberg-selove \cite{F.Ajzenberg-Selove1991}, Meyer \cite{H.Meyer1976}, Descouvemont \cite{P.Descouvemont2010}. Ref.~\cite{P.Descouvemont2010} presented three results, using different channel radii (4, 5, and 6 fm), labeled as $a$, $b$, and $c$.}\label{fig4}
\end{figure}

The last column in Table~\ref{tab1} shows the average accuracies and errors of the trained model on the test dataset. The results show that the method and current model are quite effective. We also tested the model performance using the values of eliminated parameters used to calculate the cross sections during the training data generation. Fig.~\ref{fig2} shows accuracies and normalized errors of the predicted parameters for cross sections calculated with different values of eliminated parameters. The model performances have little dependence -- less than $\sim$2$\%$ accuracy and $\sim$1 keV (35$\%$) error variations for classification and regression, respectively -- on these parameters. This demonstrates that the deep learning model performs well on data with any values of the bypassed parameters. By using more training data or fine-tuning model hyperparameters, this performance can be further improved if needed. 

\begin{table*}[!t]
\resizebox{0.7\textwidth}{!}{
\def\arraystretch{0.5}
\begin{threeparttable}
\caption{Predictions of $E$ and $\Gamma$ for the measurement data along with the previous results. For the current model, the median values of the distributions are presented with the standard deviations as representations. Ref. \cite{R.Azuma2010} also used the capture reaction data in the $R$-matrix fitting.}\label{tab2}%
\begin{tabular}{ccccccccccccc}
\toprule
Parameters && Current Model && Ref. \cite{H.Meyer1976}\tnote{1} && Ref. \cite{R.Azuma2010} && Ref. \cite{P.Descouvemont2010}\tnote{1} && Ref. \cite{P.Descouvemont2010}\tnote{1} && Ref. \cite{P.Descouvemont2010}\tnote{1} \\
\midrule
$E_{1}$ (MeV) && 0.4244(2) && 0.424 && 0.426(3) && 0.427 && 0.427 && 0.427 \\ \\
$E_{2}$ (MeV) && 1.5590(12) && 1.558 && 1.556(1) && 1.560 && 1.559 && 1.558 \\ \\
$E_{3}$ (MeV) && 1.6029(8) && 1.604 && 1.602(2) && 1.603 && 1.604 && 1.606 \\ \\
$\Gamma_{1}$ (keV) && 34.3(3) && 33 && 34.1(8) && 33.8 && 32.9 && 30.9  \\ \\
$\Gamma_{2}$ (keV) && 55.0(8) && 55 && 57.9(17) && 51.4 && 51.4 && 51.3 \\ \\
$\Gamma_{3}$ (keV) && 50.3(7) && 50 && 48.3(19) && 48.1 && 48.1 && 47.8 \\ \\
$a_c$ (fm) && - && 4.0 && 3.4 && 4.0 && 5.0 && 6.0 \\
\\ Background Poles && - && $3/2^{+}$ state\tnote{2} && \tnote{3} && \tnote{3} && \tnote{3} && \tnote{3} \\
\bottomrule
\end{tabular}
\begin{tablenotes}
 \item [1] The uncertainties are not presented in the papers.
 \item [2] While it was not explicitly stated as a background pole, the experimentally known $3/2^{+}$ level at $E=5.860$ MeV with $\Gamma=1400$ keV was included in the calculation in ref.~\cite{H.Meyer1976}. 
 \item [3] No background pole was considered.
\end{tablenotes}
\end{threeparttable}}
\end{table*}

Fig.~\ref{fig3} shows a calibration of the prediction uncertainty. For discrete parameters $J$ and $\pi$, the confidence level is the predicted probability from the softmax function, and the accuracy is the fraction of correct predictions. For continuous parameters $E$ and $\Gamma$, the confidence levels are calculated from their predicted ranges surrounding their median values, and the accuracies are the fraction of true data points included in the corresponding ranges. The dashed black line shows the ideal case where the confidence level is equal to the accuracy. It is clear that our probabilistic model achieves a performance close to this ideal case.

We derived the resonance parameters from the target measurement data of ref. \cite{H.Meyer1976} using the model. Fig.~\ref{fig4} and Table~\ref{tab2} show the model predictions on the measurement data, along with the results of previously published $R$-matrix analyses to extract $E$ and $\Gamma$ \cite{H.Meyer1976, R.Azuma2010, P.Descouvemont2010}.  Our predictions of the spin-parities of second and third states are, at $\sim$100$\%$ probability, $J_2=3/2$, $\pi_2=-$, and $J_3=5/2$, in agreement with the literature \cite{F.Ajzenberg-Selove1991}. The predictions of $E$ and $\Gamma$ are for the most part consistent with those of the previous studies, demonstrating the effectiveness of our model. Furthermore, the previously analyses used different channel radius values and background pole configurations which influenced their results, whereas our method did not require any use of these extraneous parameters.

The proposed method can also be utilized for interpolations or extrapolations of measured data by properly configuring the outputs of the deep learning model. Generalized models that address numerous similar problems can also be built, as well as pre-trained models that only need to be slightly adjusted for a specific task \cite{J.Devlin2019, F.Zhuang2021}. High capacity neural networks with sufficient depth can learn patterns in training data from $\it{any}$ physics model. Additionally, data preparation is relatively effortless compared to typical deep learning applications that require piecemeal hand-labeling, significantly easing the model training process. 

For cases where the values of the eliminated parameters in the physics model are unknown, the performance of the deep learning model -- that is, its ability to accurately predict observable parameters -- can exceed that of the physics model. Alternatively, if they could (by some means) be correctly specified, then the performances of the deep learning model and physics model are comparable. The deep learning model can accurately identify distinctive features in the data, features that arise from variations of labels in training data. In cases where variations of eliminated parameters can impact the features, the deep learning model predictions may have high uncertainties with diminished performance \cite{A.Kiureghian2009, A.Kendall2017, E.Hullermeier2021}. 

While it is challenging to make unambiguous quantitative one-on-one comparisons with conventional optimization methods, there are a number of clear advantages of our probabilistic neural network-based method. First, once a deep learning model is trained, the prediction calculation is normally very fast, with negligible computational times \cite{A.Chua2020, H.Gabbard2022}. Second, making probabilistic predictions about discrete quantities is straightforward using a classification approach, either with softmax functions or by defining specific distributions suitable for the likelihoods of the quantities. Third, the confidence level of the uncertainties in the model predictions can be rigorously evaluated through an analysis of a subset of the data (i.e., a test dataset), using well-developed measures such as a reliability diagram to assess the quality of the uncertainties \cite{A.Dawid1982, C.Guo2017, Y.Ovadia2019, L.Jospin2022}. Robust methods also exist for cases where the uncertainty is not properly calibrated \cite{C.Guo2017, V.Kuleshov2018, Y.Ovadia2019}. 

In this study, we introduce a method to analyze measurement data using deep probabilistic neural networks that offers numerous advantages over traditional approaches with phenomenological models. Introduction of deep ensembles combines the abilities of neural networks to generate solutions of complex phenomena with an easy-to-use, statistically robust mechanism to quantify uncertainties of model predictions. We also showed that we can build a deep learning model trained to perform accurate predictions from data without specific determination of a subset of phenomenological model parameters. Additionally, we successfully utilized the Transformer in a physics task with some suitable revisions. We anticipate that this approach can be widely used for data analyses in a variety of fields and to boost developments of reliable deep learning applications.

\section*{Acknowledgements}

We thank D. Phillips for helpful discussions on this study. This work was supported by the National Research Foundation of Korea (NRF) grants funded by the Korea government (MSIT) (Grants No. 2016R1A5A1013277 and No. 2020R1A2C1005981). This work was also supported in part by the Institute for Basic Science (Grant No. IBS-R031-D1), by the National Science Foundation (Grant No. NSF PHY-2011890) and by the U.S. Dept. of Energy (DOE), Office of Science, Office of Nuclear Physics under contract DE-AC05-00OR22725. Computational works for this research were performed on the data analysis hub, Olaf in the IBS Research Solution Center.


\begin{thebibliography}{10}
\expandafter\ifx\csname url\endcsname\relax
  \def\url#1{\texttt{#1}}\fi
\expandafter\ifx\csname urlprefix\endcsname\relax\def\urlprefix{URL }\fi
\expandafter\ifx\csname href\endcsname\relax
  \def\href#1#2{#2} \def\path#1{#1}\fi

\bibitem{C.Cutler1994}
C.~{Cutler}, {\'E}.~E. {Flanagan}, {Gravitational waves from merging compact
  binaries: How accurately can one extract the binary's parameters from the
  inspiral waveform?}, Physical Review D 49~(6) (1994)
  2658--2697.
\newblock \href {http://arxiv.org/abs/gr-qc/9402014}
  {\path{arXiv:gr-qc/9402014}}, \href
  {https://doi.org/10.1103/PhysRevD.49.2658}
  {\path{doi:10.1103/PhysRevD.49.2658}}.

\bibitem{P.Moller2016}
P.~{M{\"o}ller}, A.~J. {Sierk}, T.~{Ichikawa}, H.~{Sagawa}, {Nuclear
  ground-state masses and deformations: FRDM(2012)}, Atomic Data and Nuclear
  Data Tables 109 (2016) 1--204.
\newblock \href {http://arxiv.org/abs/1508.06294} {\path{arXiv:1508.06294}},
  \href {https://doi.org/10.1016/j.adt.2015.10.002}
  {\path{doi:10.1016/j.adt.2015.10.002}}.

\bibitem{H.Wiltsche2020}
H.~A. Wiltsche, P.~Berghofer, Phenomenological approaches to physics, Springer,
  2020.

\bibitem{D.Everett2021}
D.~{Everett}, W.~{Ke}, J.~F. {Paquet}, G.~{Vujanovic}, S.~A. {Bass}, L.~{Du},
  C.~{Gale}, M.~{Heffernan}, U.~{Heinz}, D.~{Liyanage}, M.~{Luzum},
  A.~{Majumder}, M.~{McNelis}, C.~{Shen}, Y.~{Xu}, A.~{Angerami}, S.~{Cao},
  Y.~{Chen}, J.~{Coleman}, L.~{Cunqueiro}, T.~{Dai}, R.~{Ehlers}, H.~{Elfner},
  W.~{Fan}, R.~J. {Fries}, F.~{Garza}, Y.~{He}, B.~V. {Jacak}, P.~M. {Jacobs},
  S.~{Jeon}, B.~{Kim}, M.~{Kordell}, A.~{Kumar}, S.~{Mak}, J.~{Mulligan},
  C.~{Nattrass}, D.~{Oliinychenko}, C.~{Park}, J.~H. {Putschke}, G.~{Roland},
  B.~{Schenke}, L.~{Schwiebert}, A.~{Silva}, C.~{Sirimanna}, R.~A. {Soltz},
  Y.~{Tachibana}, X.~N. {Wang}, R.~L. {Wolpert}, {JETSCAPE Collaboration},
  {Phenomenological Constraints on the Transport Properties of QCD Matter with
  Data-Driven Model Averaging}, Physical Review Letters 126~(24) (2021) 242301.
\newblock \href {http://arxiv.org/abs/2010.03928} {\path{arXiv:2010.03928}},
  \href {https://doi.org/10.1103/PhysRevLett.126.242301}
  {\path{doi:10.1103/PhysRevLett.126.242301}}.

\bibitem{R.DAgostino1986}
R.~B. D'Agostino, M.~A. Stephens, Goodness-of-Fit Techniques, Marcel Dekker,
  Inc., USA, 1986.

\bibitem{U.Toussaint2011}
U.~{von Toussaint}, {Bayesian inference in physics}, Reviews of Modern Physics
  83~(3) (2011) 943--999.
\newblock \href {https://doi.org/10.1103/RevModPhys.83.943}
  {\path{doi:10.1103/RevModPhys.83.943}}.

\bibitem{A.Dawid1982}
A.~P. {Dawid},
  {The
  well-calibrated bayesian}, Journal of the American Statistical Association
  77~(379) (1982) 605--610.

\bibitem{D.Smith2007}
D.~L. Smith, S.~A. Badikov, E.~V. Gai, S.-Y. Oh, T.~Kawano, N.~M. Larson, V.~G.
  Pronyaev,
 {Perspectives
  on Peelle's pertinent puzzle}, IAEA, International Atomic Energy Agency
  (IAEA), 2007.

\bibitem{P.Descouvemont2010}
P.~{Descouvemont}, D.~{Baye}, {The R-matrix theory}, Reports on Progress in
  Physics 73~(3) (2010) 036301.
\newblock \href {http://arxiv.org/abs/1001.0678} {\path{arXiv:1001.0678}},
  \href {https://doi.org/10.1088/0034-4885/73/3/036301}
  {\path{doi:10.1088/0034-4885/73/3/036301}}.

\bibitem{R.deBoer2017}
R.~J. {deBoer}, J.~{G{\"o}rres}, M.~{Wiescher}, R.~E. {Azuma}, A.~{Best}, C.~R.
  {Brune}, C.~E. {Fields}, S.~{Jones}, M.~{Pignatari}, D.~{Sayre}, K.~{Smith},
  F.~X. {Timmes}, E.~{Uberseder}, {The $^{12}$C({\ensuremath{\alpha}}
  ,{\ensuremath{\gamma}} )$^{16}$O reaction and its implications for stellar
  helium burning}, Reviews of Modern Physics 89~(3) (2017) 035007.
\newblock \href {http://arxiv.org/abs/1709.03144} {\path{arXiv:1709.03144}},
  \href {https://doi.org/10.1103/RevModPhys.89.035007}
  {\path{doi:10.1103/RevModPhys.89.035007}}.

\bibitem{L.Wright2022}
L.~G. {Wright}, T.~{Onodera}, M.~M. {Stein}, T.~{Wang}, D.~T. {Schachter},
  Z.~{Hu}, P.~L. {McMahon}, {Deep physical neural networks trained with
  backpropagation}, Nature 601~(7894) (2022) 549--555.
\newblock \href {https://doi.org/10.1038/s41586-021-04223-6}
  {\path{doi:10.1038/s41586-021-04223-6}}.

\bibitem{F.Ashtiani2022}
F.~{Ashtiani}, A.~J. {Geers}, F.~{Aflatouni}, {An on-chip photonic deep neural
  network for image classification}, Nature 606~(7914) (2022) 501--506.
\newblock \href {http://arxiv.org/abs/2106.11747} {\path{arXiv:2106.11747}},
  \href {https://doi.org/10.1038/s41586-022-04714-0}
  {\path{doi:10.1038/s41586-022-04714-0}}.

\bibitem{J.Jumper2021}
J.~{Jumper}, R.~{Evans}, A.~{Pritzel}, T.~{Green}, M.~{Figurnov},
  O.~{Ronneberger}, K.~{Tunyasuvunakool}, R.~{Bates}, A.~{{\v{Z}}{\'\i}dek},
  A.~{Potapenko}, A.~{Bridgland}, C.~{Meyer}, S.~A.~A. {Kohl}, A.~J. {Ballard},
  A.~{Cowie}, B.~{Romera-Paredes}, S.~{Nikolov}, R.~{Jain}, J.~{Adler},
  T.~{Back}, S.~{Petersen}, D.~{Reiman}, E.~{Clancy}, M.~{Zielinski},
  M.~{Steinegger}, M.~{Pacholska}, T.~{Berghammer}, S.~{Bodenstein},
  D.~{Silver}, O.~{Vinyals}, A.~W. {Senior}, K.~{Kavukcuoglu}, P.~{Kohli},
  D.~{Hassabis}, {Highly accurate protein structure prediction with AlphaFold},
  Nature 596~(7873) (2021) 583--589.
\newblock \href {https://doi.org/10.1038/s41586-021-03819-2}
  {\path{doi:10.1038/s41586-021-03819-2}}.

\bibitem{M.Raissi2019}
M.~{Raissi}, P.~{Perdikaris}, G.~E. {Karniadakis}, {Physics-informed neural
  networks: A deep learning framework for solving forward and inverse problems
  involving nonlinear partial differential equations}, Journal of Computational
  Physics 378 (2019) 686--707.
\newblock \href {https://doi.org/10.1016/j.jcp.2018.10.045}
  {\path{doi:10.1016/j.jcp.2018.10.045}}.

\bibitem{C.Kim2023}
C.~H. Kim, S.~Ahn, K.~Y. Chae, J.~Hooker, G.~V. Rogachev,
  {Noise
  signal identification in time projection chamber data using deep learning
  model}, Nuclear Instruments and Methods in Physics Research Section A:
  Accelerators, Spectrometers, Detectors and Associated Equipment 1048 (2023)
  168025.
\newblock \href {https://doi.org/https://doi.org/10.1016/j.nima.2023.168025}
  {\path{doi:https://doi.org/10.1016/j.nima.2023.168025}}.

\bibitem{C.Kim2023-2}
C.~H. {Kim}, S.~{Ahn}, K.~Y. {Chae}, J.~{Hooker}, G.~V. {Rogachev}, {Restoring
  Original Signal From Pile-up Signal using Deep Learning}, arXiv e-prints
  (2023) arXiv:2304.14496\href {http://arxiv.org/abs/2304.14496}
  {\path{arXiv:2304.14496}}, \href {https://doi.org/10.48550/arXiv.2304.14496}
  {\path{doi:10.48550/arXiv.2304.14496}}.

\bibitem{H.Gabbard2022}
H.~{Gabbard}, C.~{Messenger}, I.~S. {Heng}, F.~{Tonolini}, R.~{Murray-Smith},
  {Bayesian parameter estimation using conditional variational autoencoders for
  gravitational-wave astronomy}, Nature Physics 18~(1) (2022) 112--117.
\newblock \href {http://arxiv.org/abs/1909.06296} {\path{arXiv:1909.06296}},
  \href {https://doi.org/10.1038/s41567-021-01425-7}
  {\path{doi:10.1038/s41567-021-01425-7}}.

\bibitem{A.Boehnlein2022}
A.~Boehnlein, M.~Diefenthaler, N.~Sato, M.~Schram, V.~Ziegler, C.~Fanelli,
  M.~Hjorth-Jensen, T.~Horn, M.~P. Kuchera, D.~Lee, W.~Nazarewicz,
  P.~Ostroumov, K.~Orginos, A.~Poon, X.-N. Wang, A.~Scheinker, M.~S. Smith,
  L.-G. Pang,
  {Colloquium:
  Machine learning in nuclear physics}, Rev. Mod. Phys. 94 (2022) 031003.
\newblock \href {https://doi.org/10.1103/RevModPhys.94.031003}
  {\path{doi:10.1103/RevModPhys.94.031003}}.

\bibitem{M.Abdar2021}
M.~Abdar, F.~Pourpanah, S.~Hussain, D.~Rezazadegan, L.~Liu, M.~Ghavamzadeh,
  P.~Fieguth, X.~Cao, A.~Khosravi, U.~R. Acharya, V.~Makarenkov, S.~Nahavandi,
 {A
  review of uncertainty quantification in deep learning: Techniques,
  applications and challenges}, Information Fusion 76 (2021) 243--297.
\newblock \href {https://doi.org/https://doi.org/10.1016/j.inffus.2021.05.008}
  {\path{doi:https://doi.org/10.1016/j.inffus.2021.05.008}}.

\bibitem{L.Jospin2022}
L.~V. Jospin, H.~Laga, F.~Boussaid, W.~Buntine, M.~Bennamoun, Hands-on bayesian
  neural networks—a tutorial for deep learning users, IEEE Computational
  Intelligence Magazine 17~(2) (2022) 29--48.
\newblock \href {https://doi.org/10.1109/MCI.2022.3155327}
  {\path{doi:10.1109/MCI.2022.3155327}}.

\bibitem{K.Murphy2022}
K.~P. Murphy, Probabilistic machine learning: an introduction, MIT press, 2022.

\bibitem{B.Lakshminarayanan2017}
B.~Lakshminarayanan, A.~Pritzel, C.~Blundell,
  {Simple
  and scalable predictive uncertainty estimation using deep ensembles}, in:
  I.~Guyon, U.~V. Luxburg, S.~Bengio, H.~Wallach, R.~Fergus, S.~Vishwanathan,
  R.~Garnett (Eds.), Advances in Neural Information Processing Systems,
  Vol.~30, Curran Associates, Inc., 2017.

\bibitem{Y.Ovadia2019}
Y.~Ovadia, E.~Fertig, J.~Ren, Z.~Nado, D.~Sculley, S.~Nowozin, J.~Dillon,
  B.~Lakshminarayanan, J.~Snoek,
  {Can
  you trust your model\textquotesingle s uncertainty? evaluating predictive
  uncertainty under dataset shift}, in: H.~Wallach, H.~Larochelle,
  A.~Beygelzimer, F.~d\textquotesingle Alch\'{e}-Buc, E.~Fox, R.~Garnett
  (Eds.), Advances in Neural Information Processing Systems, Vol.~32, Curran
  Associates, Inc., 2019.

\bibitem{A.Wilson2020}
A.~G. Wilson, P.~Izmailov,
  {Bayesian
  deep learning and a probabilistic perspective of generalization}, in:
  H.~Larochelle, M.~Ranzato, R.~Hadsell, M.~Balcan, H.~Lin (Eds.), Advances in
  Neural Information Processing Systems, Vol.~33, Curran Associates, Inc.,
  2020, pp. 4697--4708.

\bibitem{S.Artemov2003}
S.~Artemov, E.~Zaparov, G.~Nie, Asymptotic normalization factors for light
  nuclei from proton transfer reactions, Bulletin of the Russian Academy of
  Sciences-Physics 67~(11) (2003) 1741--1746.

\bibitem{R.Azuma2010}
R.~E. {Azuma}, E.~{Uberseder}, E.~C. {Simpson}, C.~R. {Brune}, H.~{Costantini},
  R.~J. {de Boer}, J.~{G{\"o}rres}, M.~{Heil}, P.~J. {Leblanc}, C.~{Ugalde},
  M.~{Wiescher}, {AZURE: An R-matrix code for nuclear astrophysics}, Physical
  Review C 81~(4) (2010) 045805.
\newblock \href {https://doi.org/10.1103/PhysRevC.81.045805}
  {\path{doi:10.1103/PhysRevC.81.045805}}.

\bibitem{A.Lane1958}
A.~M. {Lane}, R.~G. {Thomas}, {R-Matrix Theory of Nuclear Reactions}, Reviews
  of Modern Physics 30~(2) (1958) 257--353.
\newblock \href {https://doi.org/10.1103/RevModPhys.30.257}
  {\path{doi:10.1103/RevModPhys.30.257}}.

\bibitem{H.Fynbo2005}
H.~O.~U. {Fynbo}, C.~A. {Diget}, U.~C. {Bergmann}, M.~J.~G. {Borge},
  J.~{Cederk{\"a}ll}, P.~{Dendooven}, L.~M. {Fraile}, S.~{Franchoo}, V.~N.
  {Fedosseev}, B.~R. {Fulton}, W.~{Huang}, J.~{Huikari}, H.~B. {Jeppesen},
  A.~S. {Jokinen}, P.~{Jones}, B.~{Jonson}, U.~{K{\"o}ster}, K.~{Langanke},
  M.~{Meister}, T.~{Nilsson}, G.~{Nyman}, Y.~{Prezado}, K.~{Riisager},
  S.~{Rinta-Antila}, O.~{Tengblad}, M.~{Turrion}, Y.~{Wang}, L.~{Weissman},
  K.~{Wilhelmsen}, J.~{{\"A}yst{\"o}}, {ISOLDE Collaboration}, {Revised rates
  for the stellar triple-{\ensuremath{\alpha}} process from measurement of
  $^{12}$C nuclear resonances}, Nature 433~(7022) (2005) 136--139.

\bibitem{E.Adelberger2011}
E.~G. {Adelberger}, A.~{Garc{\'\i}a}, R.~G.~H. {Robertson}, K.~A. {Snover},
  A.~B. {Balantekin}, K.~{Heeger}, M.~J. {Ramsey-Musolf}, D.~{Bemmerer},
  A.~{Junghans}, C.~A. {Bertulani}, J.~W. {Chen}, H.~{Costantini}, P.~{Prati},
  M.~{Couder}, E.~{Uberseder}, M.~{Wiescher}, R.~{Cyburt}, B.~{Davids}, S.~J.
  {Freedman}, M.~{Gai}, D.~{Gazit}, L.~{Gialanella}, G.~{Imbriani},
  U.~{Greife}, M.~{Hass}, W.~C. {Haxton}, T.~{Itahashi}, K.~{Kubodera},
  K.~{Langanke}, D.~{Leitner}, M.~{Leitner}, P.~{Vetter}, L.~{Winslow}, L.~E.
  {Marcucci}, T.~{Motobayashi}, A.~{Mukhamedzhanov}, R.~E. {Tribble}, K.~M.
  {Nollett}, F.~M. {Nunes}, T.~S. {Park}, P.~D. {Parker}, R.~{Schiavilla},
  E.~C. {Simpson}, C.~{Spitaleri}, F.~{Strieder}, H.~P. {Trautvetter},
  K.~{Suemmerer}, S.~{Typel}, {Solar fusion cross sections. II. The pp chain
  and CNO cycles}, Reviews of Modern Physics 83~(1) (2011) 195--246.
\newblock \href {http://arxiv.org/abs/1004.2318} {\path{arXiv:1004.2318}},
  \href {https://doi.org/10.1103/RevModPhys.83.195}
  {\path{doi:10.1103/RevModPhys.83.195}}.

\bibitem{A.Tumino2018}
A.~{Tumino}, C.~{Spitaleri}, M.~{La Cognata}, S.~{Cherubini}, G.~L. {Guardo},
  M.~{Gulino}, S.~{Hayakawa}, I.~{Indelicato}, L.~{Lamia}, H.~{Petrascu}, R.~G.
  {Pizzone}, S.~M.~R. {Puglia}, G.~G. {Rapisarda}, S.~{Romano}, M.~L. {Sergi},
  R.~{Spart{\'a}}, L.~{Trache}, {An increase in the $^{12}$C + $^{12}$C fusion
  rate from resonances at astrophysical energies}, Nature 557~(7707) (2018)
  687--690.
\newblock \href {https://doi.org/10.1038/s41586-018-0149-4}
  {\path{doi:10.1038/s41586-018-0149-4}}.

\bibitem{J.Bishop2022}
J.~{Bishop}, C.~E. {Parker}, G.~V. {Rogachev}, S.~{Ahn}, E.~{Koshchiy},
  K.~{Brandenburg}, C.~R. {Brune}, R.~J. {Charity}, J.~{Derkin}, N.~{Dronchi},
  G.~{Hamad}, Y.~{Jones-Alberty}, T.~{Kokalova}, T.~N. {Massey}, Z.~{Meisel},
  E.~V. {Ohstrom}, S.~N. {Paneru}, E.~C. {Pollacco}, M.~{Saxena}, N.~{Singh},
  R.~{Smith}, L.~G. {Sobotka}, D.~{Soltesz}, S.~K. {Subedi}, A.~V. {Voinov},
  J.~{Warren}, C.~{Wheldon}, {Neutron-upscattering enhancement of the
  triple-alpha process}, Nature Communications 13 (2022) 2151.
\newblock \href {https://doi.org/10.1038/s41467-022-29848-7}
  {\path{doi:10.1038/s41467-022-29848-7}}.

\bibitem{H.Meyer1976}
H.~O. {Meyer}, G.~R. {Plattner}, I.~{Sick}, {Elastic p+$^{12}$C scattering
  between 0.3 and 2 MeV}, Zeitschrift fur Physik A Hadrons and Nuclei 279~(1)
  (1976) 41--45.
\newblock \href {https://doi.org/10.1007/BF01409090}
  {\path{doi:10.1007/BF01409090}}.

\bibitem{A.Vaswani2017}
A.~Vaswani, N.~Shazeer, N.~Parmar, J.~Uszkoreit, L.~Jones, A.~N. Gomez, L.~u.
  Kaiser, I.~Polosukhin,
 {Attention
  is all you need}, in: I.~Guyon, U.~V. Luxburg, S.~Bengio, H.~Wallach,
  R.~Fergus, S.~Vishwanathan, R.~Garnett (Eds.), Advances in Neural Information
  Processing Systems, Vol.~30, Curran Associates, Inc., 2017.

\bibitem{J.Devlin2019}
J.~Devlin, M.-W. Chang, K.~Lee, K.~Toutanova,
  {{BERT}: Pre-training of deep
  bidirectional transformers for language understanding}, in: Proceedings of
  the 2019 Conference of the North {A}merican Chapter of the Association for
  Computational Linguistics: Human Language Technologies, Volume 1 (Long and
  Short Papers), Association for Computational Linguistics, Minneapolis,
  Minnesota, 2019, pp. 4171--4186.
\newblock \href {https://doi.org/10.18653/v1/N19-1423}
  {\path{doi:10.18653/v1/N19-1423}}.

\bibitem{S.Khan2022}
S.~Khan, M.~Naseer, M.~Hayat, S.~W. Zamir, F.~S. Khan, M.~Shah,
  {Transformers in vision: A survey}, ACM
  Comput. Surv. 54~(10s) (sep 2022).
\newblock \href {https://doi.org/10.1145/3505244} {\path{doi:10.1145/3505244}}.

\bibitem{H.Zhou2021}
H.~Zhou, S.~Zhang, J.~Peng, S.~Zhang, J.~Li, H.~Xiong, W.~Zhang,
  {Informer:
  Beyond efficient transformer for long sequence time-series forecasting},
  Proceedings of the AAAI Conference on Artificial Intelligence 35~(12) (2021)
  11106--11115.
\newblock \href {https://doi.org/10.1609/aaai.v35i12.17325}
  {\path{doi:10.1609/aaai.v35i12.17325}}.

\bibitem{F.Ajzenberg-Selove1991}
F.~Ajzenberg-Selove,
  {Energy
  levels of light nuclei a = 13–15}, Nuclear Physics A 523~(1) (1991) 1--196.
\newblock \href {https://doi.org/https://doi.org/10.1016/0375-9474(91)90446-D}
  {\path{doi:https://doi.org/10.1016/0375-9474(91)90446-D}}.

\bibitem{F.Zhuang2021}
F.~Zhuang, Z.~Qi, K.~Duan, D.~Xi, Y.~Zhu, H.~Zhu, H.~Xiong, Q.~He, A
  comprehensive survey on transfer learning, Proceedings of the IEEE 109~(1)
  (2021) 43--76.
\newblock \href {https://doi.org/10.1109/JPROC.2020.3004555}
  {\path{doi:10.1109/JPROC.2020.3004555}}.

\bibitem{A.Kiureghian2009}
A.~Der~Kiureghian, O.~Ditlevsen, Aleatory or epistemic? does it matter?,
  Structural safety 31~(2) (2009) 105--112.

\bibitem{A.Kendall2017}
A.~Kendall, Y.~Gal,
  {What
  uncertainties do we need in bayesian deep learning for computer vision?}, in:
  I.~Guyon, U.~V. Luxburg, S.~Bengio, H.~Wallach, R.~Fergus, S.~Vishwanathan,
  R.~Garnett (Eds.), Advances in Neural Information Processing Systems,
  Vol.~30, Curran Associates, Inc., 2017.

\bibitem{E.Hullermeier2021}
E.~H{\"u}llermeier, W.~Waegeman, Aleatoric and epistemic uncertainty in machine
  learning: An introduction to concepts and methods, Machine Learning 110
  (2021) 457--506.
\newblock \href {https://doi.org/10.1007/s10994-021-05946-3}
  {\path{doi:10.1007/s10994-021-05946-3}}.

\bibitem{A.Chua2020}
A.~J.~K. {Chua}, M.~{Vallisneri}, {Learning Bayesian Posteriors with Neural
  Networks for Gravitational-Wave Inference}, Physical Review Letters 124~(4)
  (2020) 041102.
\newblock \href {http://arxiv.org/abs/1909.05966} {\path{arXiv:1909.05966}},
  \href {https://doi.org/10.1103/PhysRevLett.124.041102}
  {\path{doi:10.1103/PhysRevLett.124.041102}}.

\bibitem{C.Guo2017}
C.~Guo, G.~Pleiss, Y.~Sun, K.~Q. Weinberger,
  {On calibration of modern
  neural networks}, in: D.~Precup, Y.~W. Teh (Eds.), Proceedings of the 34th
  International Conference on Machine Learning, Vol.~70 of Proceedings of
  Machine Learning Research, PMLR, 2017, pp. 1321--1330.

\bibitem{V.Kuleshov2018}
V.~Kuleshov, N.~Fenner, S.~Ermon,
 {Accurate
  uncertainties for deep learning using calibrated regression}, in: J.~Dy,
  A.~Krause (Eds.), Proceedings of the 35th International Conference on Machine
  Learning, Vol.~80 of Proceedings of Machine Learning Research, PMLR, 2018,
  pp. 2796--2804.

\end{thebibliography}



\end{document}